# STANDARDIZATION OF INFORMATION SYSTEMS DEVELOPMENT PROCESSES AND BANKING INDUSTRY ADAPTATIONS

Zuhal Tanrikulu[1] and Tuna Ozcer[2]

[1] Department of Management Information Systems, Bogazici University, Istanbul, Turkey
`tanrikul@boun.edu.tr`

[2] Finans Bank, Istanbul, Turkey
`tuna.ozcer@finansbank.com.tr`

## ABSTRACT

*This paper examines the current system development processes of three major Turkish banks in terms of compliance to internationally accepted system development and software engineering standards to determine the common process problems of banks. After an in-depth investigation into system development and software engineering standards, related process-based standards were selected. Questions were then prepared covering the whole system development process by applying the classical Waterfall life cycle model. Each question is made up of guidance and suggestions from the international system development standards. To collect data, people from the information technology departments of three major banks in Turkey were interviewed. Results have been aggregated by examining the current process status of the three banks together. Problematic issues were identified using the international system development standards.*

## 1. INTRODUCTION

The business environment is becoming more technologically focused. Current business processes rely heavily on information systems within industries. Complexity and the increasing numbers of information systems force companies to establish processes to perform business functions on information systems and operate in a more controlled environment. Xia and Lee [1] proposed to define four components of information systems development project complexity: structural organizational complexity, structural Information Technology (IT) complexity, dynamic organizational complexity, and dynamic IT complexity.

In addition to the necessity of processes related to information systems, reports published by several companies indicate a high percentage of failure for information systems projects. For example, CHAOS research performed by the Standish Group [2] covering several industries, including banking, securities, manufacturing, retail, wholesale, health care, insurance, services, and local, state, and federal organizations, found that:

- 32% of all software projects are completed on time and within budget, with all functions and features as initially specified
- 44% of the projects are completed over-budget and over the time estimate, offering fewer features and functions than originally specified
- 24% of the software projects are cancelled at some point during the development life cycle

Moreover, the research has focused on discovering why software projects fail and listed 10 main reasons for project success [2], [3]:

1. User Involvement
2. Executive Management Support
3. Clear Statement of Requirements
4. Emotional Maturity
5. Optimizing Scope
6. Agile Process
7. Project Management Expertise
8. Skilled Resources
9. Execution
10. Tools and Infrastructure

When these 10 aspects for success are observed, it becomes obvious that most of the aspects are related to well-defined processes that reside somewhere in the system development process. Below are some discussions related to the reasons most related to the processes.

- User involvement in an information system development project is succeeded by several methods, such as defining the system requirements together.
- Executive management support can be ensured by assigning a business sponsor to a project.
- A clear statement of requirements can be achieved by reviewing requirement definition documents and refining customer requirements.
- Emotional maturity is related to the project manager's ability, which makes sure that the project members abide by the common purpose and effective use of ecosystems of the organization to support the project.
- Optimizing scope relates to validating customer requirements in terms of feasibility within the process.
- Project management is related to planning each detail of a project, such as resources, risks, scheduling and the following up of each plan in a timely manner throughout the process.

Ganesh and Mehta [4], in a study about Enterprise Resource Planning (ERP) systems development projects, stated that the top three Critical Failure Factors of the projects are poor quality of testing, unrealistic expectations from top management concerning the systems, and poor top management support. In a research study on systems development, Ravichandran and Rai [5] identified top management leadership, a sophisticated management infrastructure, process management efficacy, and stakeholder participation as important elements of a quality oriented organizational system for systems development. Their results suggest that software quality goals are best attained when top management creates a management infrastructure that promotes improvements in process design and encourages stakeholders to evolve the design of the development process.

Another team of researchers focused on source code internal quality evaluation using the ISO/IEC-9126 standard as a frame of reference.Their methodology for assessment was a code based on internal quality, which consists of six characteristics: functionality, concerned with what the software does to fulfill user needs; reliability, evaluating software's capability to maintain a specified level of performance; usability, assessing how understandable and usable the software is; efficiency, evaluating the capability of the software to exhibit the required performance with regards to the amount of resources needed; maintainability, concerned with the software's capability to be modified; and portability, measuring the software's capability to be transferred across environments [6].

Source codes are only one item of information systems, but they consist of several aspects. Therefore, it is hard to assess and assure the quality of information systems. As a result, it is clear that we need more metrics and standards for the assessment of the complete system.

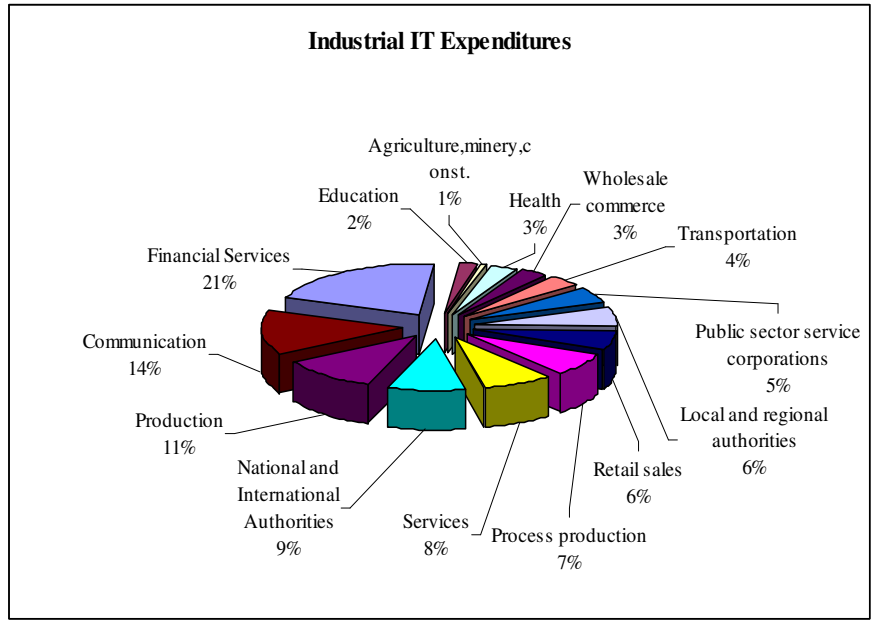

Figure 1. Industrial IT expenditures

As mentioned in the previous discussions, it is obvious that most of the success factors related to the information systems projects are process-centric and organizational. Only about 20% of a project's cost is for the software developed; the rest of the cost is in support of the project's business bureaucracy [2]. On the other hand, the organizational complexity of banks in terms of IT functions and system development efforts requires well-established processes and the proper execution of processes with predefined policies and procedures.

Today's banking industry relies heavily on information systems for most of its functions. Due to increasing customers and transactions, banking is a major industry of concern with an expanding organizational structure and intensive information systems expenditures [7]. Financial services that include banking have the highest IT expenditures among industries in the world [8]. Figure 1 is a summary of IT expenditures by industries for the year 2006, where financial services that include the banking industry have the highest investments among other industries in the world.

Although there are numerous international studies using system development and software engineering standards, during a literature review of the existing bibliography in terms of process assessment and system development in Turkish banking industry, several academic studies on the subject were identified. Kalaycı [9] performed a software process assessment of the Turkish software industry by discussing software maturity models, such as the Capability

Maturity Model Integration (CMMI), Bootstrap, Trillium, Software Technology Diagnostic, Software Process Improvement and Capability Determination (SPICE). This study classified the major sectors as package programs, services, special projects, and military projects. Firms have been identified to perform the assessment according to the major sectors. Data on process assessment has been obtained using a questionnaire extracted from the CMM maturity model at four software firms. Although their study carries out the same logical path and similar types of questions, this study does not conclude with a problem list resulting from the CMMI model.

Tarhan [10] applied the Software Best Practice Questionnaire developed by the European Software Institute (ESI) to 30 software-developing organizations in Turkey and compared the results with the implementations of the same questionnaire to European countries by the European Union. The study performs the assessment in the dimensions of software process maturity and software best practices. This study has a common issue with our study, covering financial and insurance sector companies. This study emphasizes the quantitative assessment by calculating maturity levels and best practices of the organizations and comparing the results with the European assessment performed in 1995 to compare the adoption levels of organizations by sectors.

## 2. RESEARCH QUESTIONS

The problem that this study will be touching on is the examination of current system development processes in the banking industry using references such as international and process-centric systems development and software engineering standards. For this purpose, the following research questions are discussed:

1. Most spending in the IT industry occurs in the banks managing information systems development processes compliant with the commonly accepted international standards. Hence, what is the current status and what is the expected status?
2. What is the gap between the current and expected status? What is the problem that is caused by this gap?
3. Do the banks have common problems related to standards compliance at several stages of the system development process?

## 3. METHODOLOGY

Our study has been carried out in several phases. While selecting the standards to use for the checklist preparation, the following criteria has been used.

*Correspondence* was modeled by a stage of the classical Waterfall system development model. The standard was selected if it corresponded to one of the Waterfall life cycle phases: feasibility, analysis, design, coding, testing, implementation, maintenance, and review [11], [12]. Definitions of the Waterfall model stages provided in the literature have been used for this purpose.

*Being process-centric* and standards that discuss the process based issues are preferred to the technical issues. Moreover, standards which have built an input-output mechanism between sections and processes performed were selected for this phase.

Relation to a *success factor* was determined by CHAOS research. Standards were selected if the standard relates to one of the 10 success factors found by the Standish Group in 2009. 7 out of the 10 success factors found by CHAOS research are process-centric and related to the control of system development processes somewhere in the system development life cycle.

*Accessibility* refers to whether there is a standard published by the International Organization for Standardization (ISO) and adopted by the Institute of Electrical and Electronics Engineers (IEEE). This particular accessible standard has been used for the checklist.

For *planning,* IEEE standards were selected to form the system development process questions. The first reason for this choice is the relationship of IEEE standards to ISO standards by being a liaison of the ISO joint technical committee JTC1/ subcommittee SC7. Secondly, IEEE standards are easily accessible through IEEExplore, the official research portal of IEEE. While ISO standards' adopted versions can be found on IEEExplore, ISO standards are only reachable by payment at ISO's website. IEEE standards have also been preferred for discussing more detailed system development process issues than ISO. Forty-three active IEEE standards have been scanned by reading, at this stage, to use in the preparation of system development process questions. As a result, 17 out of 43 standards have been selected and used to form the system development questions.

The following standards have shown higher correspondence to different stages of the system development process:

1. IEEE Standard 1074- Software life cycle processes [13]
2. IEEE Standard 1540- Software life cycle processes risk management [14]
3. IEEE Standard 1062- Software acquisition [15]
4. IEEE Standard 1058- Software project management plans [16]
5. IEEE Standard 1233- Developing system requirements specifications [17]
6. IEEE Standard 830- Software requirements specification [18]
7. IEEE/EIA Standard 12207.2- Software life cycle processes implementation [19]
8. IEEE/EIA Standard 12207.0- Software life cycle processes [20]
9. IEEE Standard 1061- Software quality metrics [21]
10. IEEE Standard 730- Software quality assurance [22]
11. IEEE Standard 1016- Software design description [23]
12. IEEE Standard 828- Software configuration management [24]
13. IEEE Standard 829- Software testing and documentation [25]
14. IEEE Standard 1063- Software user documentation [26]
15. ISO/IEC Standard 14764 - IEEE Standard 14764 – Software life cycle processes and maintenance [27]
16. IEEE Standard 1219- Software maintenance [28]
17. IEEE Standard 1028- Software reviews [29]

While selecting the standards, it was noted that standards that have shown the highest correspondence to stages of the system development process, namely IEEE Standard 1074, IEEE Standard 12207.0- 1996, IEEE Standard 12207.2-1997, have shown a process sequence similar to the classical Waterfall life cycle model. Consequently, questions have been grouped according to the stages of the classical Waterfall life cycle model. Each question has been generated with respect to the guidance, or process, definitions provided by selected IEEE Standards. As a result, 151 questions for the whole system development process have been generated during the question preparation phase.

According to the Banking Regulatory and Supervisory Agency (BRSA) monthly bulletin, there are 10 active domestic private commercial banks in Turkey [30]. The BRSA has provided a ranking for domestic private banks. The three domestic private banks interviewed in this study were selected from the top five domestic private banks that had the highest assets in 2010 [31]. Reasons to choose domestic private banks, rather than state banks, include that they are more technologically focused and exhibit a higher dynamism in terms of technology usage and IT strategies.

After the decision that questions have matured sufficiently, interviews were performed with three major Turkish banks by asking questions to banking professionals versed in process practices. Due to the complexity of the process and the questions, questions were divided according to the area of expertise within the banks. Each project at the bank included collecting information on profiles from project managers, software designers/ developers, business/systems analysts, risk management professionals and quality assurance professionals. Moreover, each interview with a person that had any of these profiles lasted about an hour. Open ended questions were asked of the respondents. Due to the corporate confidentiality requirements of banks, a confidentiality agreement was signed and sealed by the authors of this paper. Gathered information will only be used for academic purposes and will not be shared with third parties. All interviewees were made aware of this prior to starting the interviews.

Upon interview completion, the banks' current situation of the system development project was compared with the expected situations that come from the standards. If it existed, a problem definition was created for the existing processes.

## 4. RESULTS

After discussing the current process conditions of the banks, problems common to at least two banks for each development phase were identified with respect to IEEE system development and software engineering standards. These common problems are as follows:

Project Management Phase Problems:

1.    Managerial process plans suggested by the standard are not created completely by the banks. IEEE Standard 1058.
2.    Project management plans are not managed by a formal configuration management approach. IEEE Standard 1058.
3.    Project control plans covering metrics, reporting mechanisms, and control procedures are not created. IEEE Standard 1058.
4.    Project progress is not measured using estimated plans and actual results. IEEE Standard 1074.
5.    Technical process plans covering the development process model, technical methods, tools, and techniques are not completely created. IEEE Standard 1058.
6.    Subcontractor selection criteria are not specified in the subcontractor management plan. IEEE Standard 1058.
7.    Types of risk analysis required in the risk management process are not documented. IEEE Standard 1540.
8.    Results of the risk monitoring process are not reported to project stakeholders. IEEE Standard 1540.

Feasibility Phase Problems:

1.    Banks don't have a software acquisition strategy for acquiring off-the-shelf products. IEEE Standard 1062.

Analysis Phase Problems:

1.    A formal change process is not applied to track and control changes on SRS documents. IEEE Standard 830.

Design Phase Problems:

1.   Software reviews, tests, problem reporting and corrective actions, supplier control, records collection maintenance and retention, training, risk management, glossary, quality assurance change procedure and history sections suggested by the standard are not created within software quality assurance plans. IEEE Standard 730.
2.   Draft versions of user documentation are not prepared by the design staff. IEEE/EIA Standard 12207.0.
3.   Preliminary versions of test requirements are not prepared by the design staff. IEEE/EIA Standard 12207.0.

Coding/Package Selection Phase Problems:

1.   Coding and commenting standards and procedures are not in place. IEEE Standard 1074.
2.   Software configuration management plans are not created along the process. IEEE Standard 828.
3.   Software configuration management policy is not created to be used along the process. IEEE Standard 828.
4.   Software configuration management procedure is not created to be used along the process. IEEE Standard 828.
5.   Roles and responsibilities for technical and managerial activities of the SCM process are not documented by the banks. IEEE Standard 828.
6.   An overall, detailed release management plan, including software release management objectives, release frequency, release milestones, release media, building procedures, naming conventions, branching models, and delivery media is not prepared by the banks, as suggested by the standard. IEEE Standard 1074.
7.   Access to the software libraries and retrieval of configuration items from the software libraries are not governed by formal procedures. IEEE Standard 828.
8.   Banks have not created a standard software acquisition process. IEEE Standard 1062.
9.   Each software unit, or database development effort, is not documented along the process. IEEE/EIA Standard 12207.0.
10.  Results of unit tests are not formally documented along the process. IEEE/EIA Standard 12207.0.
11.  Integration test plans are not prepared for all projects. IEEE/EIA Standard 12207.0 and IEEE/EIA Standard 12207.2.
12.  Draft versions of user documentation are not prepared in the development process. IEEE/EIA Standard 12207.0.

Testing Phase Problems:

1.   Integration plans are not prepared for all system development projects.  IEEE Standard 1074.
2.   Problems encountered during installation to test the environment are not documented along the process. IEEE Standard 1074.
3.   Test design specification documents are not prepared to specify the test approach and methods to be used and pass/fail criteria for the software features. IEEE Standard 829.
4.   Results of tests performed are not approved by authorized personnel. IEEE Standard 829.

Implementation Phase Problems:

1.   Production environment is not operated using operating instructions or standard operational procedures. IEEE Standard 1074.

2.    Formal problem management procedures to handle problems encountered at the production environment are not created by the banks.  IEEE/EIA Standard 12207.0.

3.    Procedures related to user documentation to guide the documentation process are not prepared by the banks. IEEE/EIA Standard 12207.0.

Maintenance Phase Problems:

1.    Although the impact of change to current users is considered within the feasibility study of modification, preliminary implementation plans are not created by the banks. IEEE Standard 1219.

2.    Approval regarding the satisfactory completion of maintenance is not obtained at Bank B and Bank C. IEEE/EIA Standard 12207.0.

3.    Post-operation review process is not established to assess the impact of the change to the new environment. ISO/IEC Standard 14764 and IEEE Standard 14764.

Review Phase Problems:

1.    Installation plans, maintenance plans, software configuration management plans, and software safety plans are not subject to management reviews. IEEE Standard 1028.

2.    Technical review process is not formally executed at Bank A and Bank B. IEEE Standard 1028.

3.    Maintenance manual, system building procedures, installation procedures, and release notes are not subject to technical reviews. IEEE Standard 1028.

4.    Software user documentation, maintenance manuals, and system building procedures are not subject to internal inspections. IEEE Standard 1028.

5.    Release notes and installation procedures are not subject to internal inspections. IEEE Standard 1028.

6.    Software products are not subject to walk-through reviews. IEEE Standard 1028.

7.    Design verification is not performed by Bank A, Bank B, and Bank C. IEEE/EIA Standard 12207.2.

8.    Process verification is not performed by Bank A and Bank C.  IEEE/EIA Standard 12207.2.

The list of problems illustrates that the three major Turkish banks have common process compliance problems to standards in each phase of system development. This issue can be related to many factors and includes:

•   BRSA has commenced information systems audit regulations in 2006; banks are now in the initiation phase of the projects to reach certain software process maturity levels using Control Objectives for Information Related Technology (CobiT) and Capability Maturity Model Integration (CMMI) frameworks.

•   The number of individual problems listed is very similar, which indicate that banks are all in the initiation phase for process improvements. This was also verified by the banking professionals during the interviews.

•   The highest number of problems was identified for standards that cover the largest portion of the system development life cycle. This is extremely normal, as question numbers increased due to the coverage of standards.

•   When problems by phases are observed, it is acceptable to create the result that most problematic phases include a review, project management, implementation, and testing, affiliated with the density of questions and availability of standards for these phases.

This study has demonstrated that the three banks have common problems in the following areas:

- All managerial plans suggested by the standards, such as estimation, staff, and training plans, are not prepared by the banks.
- Banks are not preparing project control plans that should include metrics, reporting mechanisms, and control procedures.
- An overall, detailed release management plan, including software release management objectives and a release frequency is not prepared by the banks. Instead, banks choose to have specific release delivery dates.
- Access to software libraries are not governed with formally documented and accepted procedures at all banks.
- The documentation of development is not performed at the banks. This would allow for the dissemination and storage of tacit knowledge, as well as increasing the development experience of technical staff.
- Test design specification documents are not prepared to specify the test approach, methods to be used and pass/fail criteria for the software features at the banks. This would allow for the design approach to be applied for software and system testing.
- Preliminary implementation plans are not created for modifications to ensure the minimal impact of changes to the existing organization.
- The post-operation review process is not established to assess the impact of the modification to the existing environment at all banks. This allows for the earlier identification of problems.
- In terms of review, software user documentation, maintenance manuals, and system build procedures are not subject to internal inspections.
- Software products are not subject to walk-through reviews to ensure knowledge sharing and collaboration between technical staff.
- Design verification is not performed to verify that design is compliant with defined system requirements and that design is traceable from system requirements.

## 5. CONCLUSIONS

Common system development process problems of major Turkish banks were determined by applying internationally accepted system development and software engineering standards. Although the study does not include all private banks in Turkey, we assume that the results from the three major Turkish banks can be extrapolated in relation to the standard compliance status of other banks in the industry. Moreover, taking the observed problems into consideration will help banks  improve their existing system development processes and reach higher project success rates. Further studies investigating other banks are appropriate and important to enhance the industrial information base and industrial facts.

The major limitation of this study is the confidentiality requirements of the Turkish banks. As a solution to this problem, confidentiality agreements were signed with the three banks.

During the bank selection process and the preparation of the introduction, it was challenging to determine the facts and figures related to the individual IT expenditures of the banks. Regulatory bodies such as the BRSA and the Banks Association of Turkey retrieve such data by accounts from the banks. However, indicators, such as IT expenditures, IT staff, and project success rates, are not included within the publications and reports published by these organizations. Moreover, banks record this historical data, but are hesitant to share such information due to strict organizational confidentiality within the industry.

Finally, as a targeted audience, this study aims to provide significant facts about industrial process status information to IT staff of Turkish banks, independent auditing companies, and all the individuals interested in process improvement and analysis using an alternate approach rather than well-known frameworks such as CobiT and CMMI. This study can be extended to several special IT governance topics, such as change management, supplier relationship management for IT departments, and software configuration management. The literature review illustrated that there is a sufficient number of standards in the expected level of details.

**Authors**

**Zuhal Tanrikulu** is an Associate Professor of Management Information Systems at Bogazici University in Istanbul. She received her B.S. in Electronics and Communication Engineering, M.S. in Computer Engineering, and Ph.D. in Management and Organization, specializing in MIS. Her current research interests focus on the implementation of information technology in organizations, analysis and design of integrated information systems, information systems security and management, web service-based information systems, and learning management systems. She is active in teaching algorithms, programming, and information systems management courses. She has been published in Educause Quarterly, International Review on Computers and Software, Management, and the Journal of the Faculty of Education.

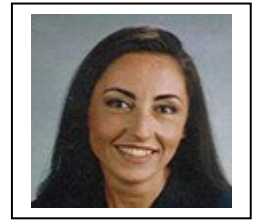

**Tuna Ozcer** works as a Process Engineer at Finansbank, Business Development and Strategy Office which helps organization to improve organizational processes and carries out projects driven by Board of Directors, CEO, and CIO. He received his B.S in Mathematical Engineering from Yıldız Technical University, Istanbul in 2004 and M.A. in Management Information Systems from Bogazici University in 2008. He is interested in Information Technology concepts including System Development, Requirements Gathering, System Analysis/Business Process Analysis.

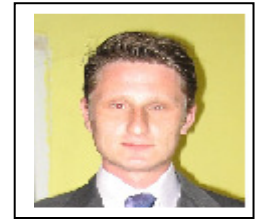